\documentclass[preprint, notitlepage]{revtex4}   
\newcommand{\be}{\begin{equation}}
\newcommand{\ee}{\end{equation}}
\newcommand{\beqn}{\begin{eqnarray}}
\newcommand{\eeqn}{\end{eqnarray}}
\def\slashi#1{\rlap{\sl/}#1}
\def\slashii#1{\setbox0=\hbox{$#1$}             
   \dimen0=\wd0                                 
   \setbox1=\hbox{\sl/} \dimen1=\wd1            
   \ifdim\dimen0>\dimen1                        
      \rlap{\hbox to \dimen0{\hfil\sl/\hfil}}   
      #1                                        
   \else                                        
      \rlap{\hbox to \dimen1{\hfil$#1$\hfil}}   
      \hbox{\sl/}                               
   \fi}                                         %
\def\slashiii#1{\setbox0=\hbox{$#1$}#1\hskip-\wd0\hbox to\wd0{\hss\sl/\/\hss}}
\begin{document}

\title{{\bf Quantum Gravity induced Lorentz invariance violation in
the Standard Model:hadrons}}

\author{Jorge Alfaro}

\affiliation{Dept. de F\'\i sica Te\'orica C-XI, Facultad de Ciencias, Univ. Aut\'onoma
de Madrid, Cantoblanco, 28049, Madrid, Spain and \\Facultad de F\'{\i}sica, Pontificia Universidad Cat\'{o}lica de Chile \\
        Casilla 306, Santiago 22, Chile.\footnote{Permanent Address.}
\\ {\tt jalfaro@puc.cl}}

\date{\today}

\begin{abstract}
The most important problem of fundamental Physics is the quantization of the gravitational field.
A main difficulty is the lack of available experimental tests that discriminate among the theories
proposed to quantize gravity. Recently we showed that the Standard Model(SM) itself contains tiny Lorentz invariance
violation(LIV) terms coming from QG. All terms depend on one arbitrary parameter $\alpha$ that set the scale
of QG effects. In this paper we obtain the LIV for mesons and nucleons and apply it to study several effects,
including the GZK anomaly. 
\end{abstract}

\maketitle

\section{Introduction}

 In recent years several proposal have been advanced to select theories and predict new phenomena
 associated to the Quantum gravitational field \cite{1,2,3,4}. Most of the new phenomenology is associated
 to some sort of Lorentz invariance violations(LIV's)\cite{5,6,bertolami}. Recently \cite{7},
 this approach has been subjected to severe criticism.

In a previous letter\cite{alfaro}, we asserted that the main effect of QG is to deform the measure of integration of Feynman graphs
at large four momenta by a tiny LIV. The classical lagrangian is unchanged.
In a similar manner, we can say that QG deforms the metric of space-time, introducing a tiny LIV
proportional to (d-4)$\alpha$, d being the dimension of space time in Dimensional Regularization
and $\alpha$ is the only arbitrary parameter in the model. Such small
LIV could be due to quantum fluctuations of the metric of space-time produced by QG:virtual black holes as
suggested in\cite{1}, D-branes as in \cite{e2}, compactification of extra-dimensions or spin-foam anisotropies
\cite{8}. A  precise derivation of $\alpha$ will have to wait for additional progress
in the available theories of QG\footnote{Such derivation must explain why the LIV
parameter is so small. Progress in this direction is in \cite{3,4,9}. There $\alpha$ appears as $(l_P/{ L})^2$,
where $l_P$ is Planck's lenght and ${ L}$ is defined by the semiclassical gravitational state
in Loop Quantum Gravity. If $L\sim 10^{11} l_P$, an $\alpha$ of the right order is obtained}

It is possible to have modified dispersion relations without a preferred frame(DSR)\cite{dsr}.
Notice, however, that in our case the classical lagrangian is invariant under usual linear Lorentz
transformations but not under DSR. So our LIV is more akin to radiative breaking of usual
Lorentz symmetry than to DSR. Moreover the regulator R defined below and the deformed metric (5) are given in a particular
inertial frame, where spatial rotational symmetry is preserved. That is why,
in this paper we are ascribing to the point of view of \cite{6} which is
widely used in the literature. The preferred frame  is the one where the Cosmic
Background Radiation is isotropic.

Within the Standard Model, such LIV implies several remarkable effects, which are wholly determined 
up to one arbitrary parameter ($\alpha$).The main effects are:

The maximal attainable velocity for
particles is not the speed of light, but depends on the specific couplings of the particles within 
the Standard Model. Noticeably, this LIV of the dispersion relations is the only acceptable,
according to the
very stringent bounds coming from the Ultra High Energy Cosmic Rays (UHECR) spectrum\cite{9}.
Moreover, the specific interactions between particles in the SM,
determine different maximum attainable velocities  for each particle, a necessary requirement to explain
the Greisen\cite{10},Zatsepin and Kuz'min\cite{11}(GZK) anomaly\cite{6,9,old}. Since the Auger\cite{auger}
experiment is expected to produce results in the near future, powerful tests
of Lorentz
invariance using the spectrum of  UHECR will be available.

Also birrefringence occurs for charged leptons, but not for gauge bosons.
In particular, photons and neutrinos have different maximum attainable
velocities. This could be tested in the next generation of neutrino detectors such as
NUBE\cite{12,13}.

Vertices in the SM will pick up a finite LIV.

This paper is organized as follows: In chapter 2, we present the LIV cutoff regulator; section 3 contains the effect
of the regulator on One Particle Irreducible Green functions(1PI); section 4 defines LIV dimensional regularization ;
Explicit one loop computations are contained in section 5; the LIV for mesons and baryons is found in chapter 6;  
Reactions thresholds  are contained in section 7; Bounds on $\alpha$ are derived in section 8; Section 9 contains 
our conclusions.

\section{Cutoff regulator}

To see what are the implications of the asymmetry in the measure for renormalizable
theories, we will mimic the Lorentz asymmetry of the measure by the replacement
$$
\int d^dk->\int d^d k R(\frac{k^2+\alpha k_0^2}{\Lambda^2})
$$
Here R is an arbitrary function, $\Lambda$
is a cutoff with mass dimensions, that will go to infinity at the end of the calculation.
We normalize $R(0)=1$ to recover the original integral. $R(\infty)=0$ to regulate the integral.
$\alpha$ is a real parameter. Notice that we are assuming that rotational invariance in space is preserved.
More general possibilities such as violation of rotational symmetry in space can be easily incorporated
in our formalism.

This regulator has the property that for logarithmically divergent integrals, the divergent term is
Lorentz invariant whereas when the cutoff goes to infinity a finite LIV part proportional
to $\alpha$ remains.

\section { One loop} 

Let D be the naive degree of divergence of a One Particle Irreducible (1PI) graph.
The change in the measure induces
modifications to the primitively log divergent integrals(D=0) In this case, the correction
amounts to a finite
LIV. The finite part of 1PI Green functions will not be affected.
Therefore, Standard Model predictions are intact, except for the maximum attainable velocity
for particles\cite{6} and interaction vertices, which receive a finite wholly determined contribution from Quantum Gravity.

Let us analyze the primitivily divergent 1PI graphs for bosons first.

{\bf Self energy:}
$\chi(p)=\chi(0)+A^{\mu\nu}p_\mu p_\nu +convergent$,
$A^{\mu\nu}=\frac{1}{2}\partial_\mu\partial_\nu\chi(0)$. We have:
$$
A^{\mu\nu}=c_2\eta^{\mu\nu}+a^{\mu\nu}
$$
$c_2$ is the log divergent wave function renormalization counterterm; $a^{\mu\nu}$
is a finite LIV. The on-shell condition is:
$$
p^2-m^2-a^{\mu\nu}p_\mu p_\nu=0
$$
If spatial rotational invariance is preserved, the nonzero components of the matrix $a$ are:
$$
a^{00}=a_0;\ \
a^{ii}=-a_1
$$
So the maximum attainable velocity for this particle will be:
\begin{equation}
c_m=\sqrt{\frac{1-a_1}{1-a_0}}\sim 1-(a_1-a_0)/2
\end{equation}

For fermions, we have the self energy graph
$$
\Sigma(p)=\Sigma(0)+s^{\mu\nu}\gamma_\nu p_\mu
$$
$s^{\mu\nu}\gamma_\nu =\partial_\mu\Sigma(0)$. Moreover
$$
s^{\mu\nu}=s\eta^{\mu\nu}+a^{\mu\nu}/2
$$
$s$ is a log divergent wave function renormalization counterterm; $a^{\mu\nu}$
is a finite LIV. The maximum attainable velocity of this particle will be given again by equation (1).

By doing explicit computations for all particles in the SM, we get definite predictions
for the LIV,
assuming a particular regulator $R$. However, the dependence on $R$ amounts to a
multiplicative factor.
So ratios of LIV's are uniquely determined.

{\bf Vertex correction}
This graph has $D=0$, so the regulator R  will induce
a tiny LIV.

{\bf Gauge Bosons}
Consider the most general quadratic
Lagrangian which is gauge invariant, but could permit LIV's \footnote{A Chern-Simons
term is absent due to the symmetry $k_\mu->-k_\mu$, which is preserved by the regulator.}
$$
L=c^{\mu\nu\alpha\beta}F_{\mu\nu}F_{\alpha\beta}
$$
$c^{\mu\nu\alpha\beta}$ is antisymmetric in $\mu\nu$ and $\alpha\beta$ and symmetric by
$(\alpha,\beta)<->(\mu,\nu)$
It implies
that the most general expression for the self-energy of the gauge boson will be

\be
\Pi^{\nu\beta}(p)=c^{\mu\nu\alpha\beta}p_\alpha p_\mu \Pi(p)
\ee

We see that
$$
p_\nu \Pi^{\nu\beta}(p)=0
$$

$c^{\mu\nu\alpha\beta}$ is given by a logarithmically divergent integral.We get:
\be
c^{\mu\nu\alpha\beta}=c_2(\eta^{\mu\alpha}\eta^{\nu\beta}-\eta^{\mu\beta}\eta^{\nu\alpha})+
a^{\mu\nu\alpha\beta}
\ee
$c_2$ is a Lorentz invariant counterterm and $a^{\mu\nu\alpha\beta}$ is a LIV.

It is clear that
the same argument applies to massive gauge bosons that got their mass by spontaneous gauge
symmetry breaking
as well as to the graviton in linearized gravity.

Explicit computations are simplified by using Dimensional
Regularization as explained below.

\section{LIV Dimensional Regularization}

We generalize dimensional regularization to a d dimensional space with an arbitrary constant
metric $g_{\mu\nu}$. We work with a positive definite metric first  and then
Wick rotate. We will illustrate the procedure with an example.
Here $g=det(g_{\mu\nu})$ and $\Delta>0$.

\beqn
\frac{1}{\sqrt{g}}\int\frac{d^dk}{(2\pi)^d}\frac{k_\mu k_\nu}{(k^2+\Delta)^n}=\nonumber\\
\frac{1}{\sqrt{g}\Gamma(n)}\int_0^\infty dt t^{n-1}\int\frac{d^dk}{(2\pi)^d}k_\mu k_\nu e^{-t(g^{\alpha\beta}k_\alpha k_\beta+\Delta)}=\nonumber\\
\frac{1}{(4\pi)^{d/2}}\frac{g_{\mu\nu}}{2}\frac{\Gamma(n-1-d/2)}{\Gamma(n)}\frac{1}{\Delta^{n-1-d/2}}
\eeqn

In the same manner, after Wick rotation, we obtain Appendix A4 of \cite{12}.

These definitions preserve gauge invariance, because the integration measure
is invariant under shifts.
To get a LIV measure, we assume that
\be
g^{\mu\nu}=\eta^{\mu\nu}+(4\pi)^2\alpha\eta^{\mu 0}\eta^{\nu 0}Res_{\epsilon=0}
\ee
where $\epsilon=2-\frac{d}{2}$ and $Res_{\epsilon=0}$ is the residue of the pole at $\epsilon=0$. A formerly divergent integral will have a pole at $\epsilon=0$, so
when we take the physical limit, $\epsilon->0$, the answer will contain a LIV term.

That is, LIV dimensional regularization consists in:

 1)Calculating the d-dimensional integrals using a general metric $g_{\mu\nu}$.

 2) Gamma matrix algebra is generalized to a general metric $g_{\mu\nu}$.

 3) At the end of the calculation, replace
 $g^{\mu\nu}=\eta^{\mu\nu}+(4\pi)^2\alpha\eta^{\mu 0}\eta^{\nu 0}Res_{\epsilon=0}$
 and then take the limit $\epsilon->0$.

To define the counterterms, we used the minimal substraction scheme(MSS); that is
we substract the poles in $\epsilon$ from the 1PI graphs.

As a concrete example, let us evaluate a typical one loop integral that appears in
the calculation of self energy graphs:
\beqn
A^{\mu\nu}=\int \frac{d^dk}{(2\pi)^d}\frac{k^\mu k^\nu}{[k^2-m^2+i0]^3}=\\
\frac{i}{(4\pi)^{d/2}}\frac{g^{\mu\nu}}{2}\frac{\Gamma(2-\frac{d}{2})}{2}\frac{1}{(m^2)^{2-\frac{d}{2}}}\\
=\frac{i}{(4\pi)^{d/2}}\frac{\eta^{\mu\nu}+(4\pi)^2\alpha\delta^\mu_0\delta^\nu_0Res_{\epsilon=0}}{2}\frac{\Gamma(2-\frac{d}{2})}{2}\frac{1}{(m^2)^{2-\frac{d}{2}}}\\
=\frac{i}{4(4\pi)^2}(\frac{\eta^{\mu\nu}}{\epsilon}+(4\pi)^2\alpha\delta^\mu_0\delta^\nu_0) +{\rm a\  finite\  LI \ term}
\eeqn

LIV Dimensional Regularization reinforces our claim that these tiny LIV's originates in
Quantum Gravity. In fact the sole change of the metric of space time is a correction of order
 $\epsilon$ to the Minkowsky metric and this is the source of the effects studied above.
Quantum Gravity is the strongest candidate to produce such effects because the gravitational
field is precisely the metric of space-time and tiny LIV modifications to the flat Minkowsky
metric may be produced by quantum fluctuations.

\section{ Explicit One loop computations} 

\subsection{Example:$g\phi^3$ in six space-time dimensions}

To illustrate the method using the cutoff regulator, we consider $g\phi^3$ in six space-time dimensions.

The one-loop contribution to the self energy of the particle is:
\be
i\chi(q)=\frac{(-gi)^2}{2}\int \frac{d^6k}{(2\pi)^6}\frac{1}{k^2-m^2+i0}\frac{1}{(k-q)^2-m^2+i0}
\ee
The term containing the LIV is:

\beqn
L(i\chi)=-2q^\mu q^\nu g^2 B_{\mu\nu}\\
B_{\mu\nu}=\int\frac{d^6k}{(2\pi)^6}\frac{k_\mu k_\nu}{(k^2-m^2+i0)^4}
\eeqn

To evaluate $B_{\mu\nu}$, introduce the regulator of the integration measure, 
\be
R=\frac{-\Lambda^2}{k^2-\Lambda^2+a k_0^2+i0}
\ee
define $k=\Lambda p$ and take the limit $\Lambda->\infty$. In this way we verify that the LIV is mass
independent. Since $a<<1$, we keep only the first order in $a$. We end up with:
\be
LB_{\mu\nu}\sim a\int \frac{d^6p}{(2\pi)^6}\frac{p_0^2 p_\mu p_\nu}{(p^2-1+i0)^2(p^2+i0)^4}
\ee
Therefore:
\be
L(i\chi)=-\frac{g^2a q_0^2 i}{24(4\pi)^3}
\ee

\subsection{LIV in the Standard Model}

We follow \cite{14,pokorski}
and use LIV Dimensional Regularization.

{\bf Photons}

In the SM the photon self-energy can be written:
\be
i\Pi^{\mu\nu}=i(q^2 g^{\mu\nu}-q^\mu q^\nu)(\frac{-23 e^2}{48\pi^2\epsilon}+finite)
\ee
so that the LIV photon self-energy in the SM is:
\beqn
L\Pi^{\mu\nu}(q)=-
\frac{23}{3}e^2\alpha q_\alpha q_\beta\nonumber\\
(\eta^{\alpha\beta}\delta^\mu_0\delta^\nu_0+\eta^{\mu\nu}\delta^\alpha_0\delta^\beta_0-\eta^{\nu\beta}\delta^\mu_0\delta^\alpha_0-\eta^{\mu\alpha}\delta^\nu_0\delta^\beta_0)
\eeqn

It follows that
the maximal attainable velocity is
\be
c_{\gamma}=1-\frac{23}{6}e^2\alpha
\ee

We have included coupling to quarks and charged leptons as well as 3 generations and color.

\subsection{Fermions}

Let us consider QED, as an example. The electron self-energy to one loop is given by:

\be
-i\Sigma_2(q)=(-i e)^2\int \frac{d^dk}{(2\pi)^d}\frac{1}{\sqrt{g}}\gamma^\mu\frac{(i\slashi k+m)}{k^2-m^2+i0}
\gamma_\mu\frac{-i}{(k-q)^2-\mu^2+i0}
\ee
To obtain the LIV, we have to evaluate (we have introduced a parameter $\Delta$ and put it to zero afterwards):
\beqn
-iL\Sigma_2(q)=2i(-ie)^2\int\frac{d^dk}{(2\pi)^d}\frac{1}{\sqrt g}\frac{\gamma^\mu \slashi{k} \gamma_\mu k.q}{(k^2-\Delta+i0)^3}\\
= -2i(-ie)^2(d-2)\int\frac{d^dk}{(2\pi)^d}\frac{1}{\sqrt g}\frac{\slashi k k.q}{(k^2-\Delta+i0)^3}\\
=(-ie)^2\frac{(d-2)}{2} \frac{1}{(4\pi)^{d/2}}\slashi q \frac{\Gamma(2-d/2)}{\Delta^{2-d/2}}
\eeqn

but
\beqn
\slashi q \frac{\Gamma(2-d/2)}{\Delta^{2-d/2}}=q_\mu (\delta^\mu _a-\frac{(4\pi)^2\alpha Res_{\epsilon=0}}{2}
\delta^\mu _0 \delta^0 _a)\gamma^a \frac{\Gamma(2-d/2)}{\Delta^{2-d/2}}=\\
\slashi q-\frac{(4\pi)^2\alpha}{2} q_0 \gamma^0
\eeqn
so,
\be
-iL\Sigma_2(q)=\frac{e^2\alpha}{2} q_0 \gamma^0
\ee

Similarly, in the SM, the fermion self-energy is given by:
\be
(4\pi^2)\Sigma(q)=-\frac{1}{\epsilon}\slashi q\sum_{graphs} (|c_V+c_A|^2 P_L+(|c_V-c_A|^2P_R) +finite
\ee
where the fermion-gauge boson vertex is:
\be
i\gamma^k(c_V-c_A\gamma^5)
\ee
and $P_L(P_R)$  are the L(R) helicity projectors. 

Therefore 
\be
L\Sigma(q)=\frac{\alpha}{2} q_0\gamma^0 \sum_{graphs} (|c_V+c_A|^2 P_L+(|c_V-c_A|^2P_R)
\ee

We apply this last result to neutrinos and charged leptons below.

{\bf Neutrinos:}
The maximal attainable velocity is
\be
c_\nu=1-(3+tan^2\theta_w)  \frac{g^2 \alpha}{8}
\ee
In this scenario, we predict that neutrinos \cite{13} emitted simultaneously
 with photons in gamma ray bursts will not arrive simultaneously to Earth . The time delay during a flight
 from a source situated at a distance
 $D$ will be of the order of $(5\times 10^{-23}) D/c\sim 5\times 10^{-6}$ s, assuming $D=10^{10}$ light-years.
 No dependence of the time delay on the energy of high energy photons or neutrinos  should be
 observed(contrast with \cite{1}). Photons will arrive earlier since $\alpha<0$(See below).
 These predictions could be tested in the next generation of neutrino detectors \cite{14}.

Using  $R_\xi$-gauges we have checked  that the LIV is gauge invariant.
The gauge parameter affects the Lorentz invariant part only.

{\bf Electron self-energy in the Weinberg-Salam model. Birrefringence:}

Define: $e_L=\frac{1-\gamma^5}{2}e$, $e_R=\frac{1+\gamma^5}{2}e$, where $e$ is the electron field. We get
\beqn
c_L=1-(\frac{g^2}{cos^2\theta_w}(sin^2\theta_w-1/2)^2 +e^2+g^2/2)\frac{\alpha}{2};\\
c_R=1-(e^2+\frac{g^2sin^4\theta_w}{cos^2\theta_w})\frac{\alpha}{2}
\eeqn
The difference in maximal speed for the left and right helicities is $\sim( 5\times 10^{-24})$.

\section{Mesons and Baryons}

In order to apply our results to the computation of the UHECR spectrum and other phenomena, we must calculate the maximal attainable
velocity of hadrons. As we mentioned before, the problem is hadronization. One way to get an estimation of the effect
is using effective lagrangians.

We use the results of \cite{ecker,fearing} for the wave function renormalization of pions and nucleons in 
the chiral lagrangian and Heavy Baryon Chiral Perturbation Theory. They get:

\beqn
Z_\pi^{-1}=1-\frac{4m_\pi^2}{3(4\pi)^2F^2}\frac{1}{\epsilon} +finite \\
Z_N^{-1}=1-\frac{9 g_A^2 m_\pi^2}{4(4\pi)^2F^2}\frac{1}{\epsilon}+finite
\eeqn

Here, $m_\pi$ is the renormalized pion mass, $F$ is the renormalized decay constant of pions and 
$g_A$ is the axial vector coupling  constant, in the chiral limit.

Using the LIV metric, we can read off the maximal attainable velocities for pions and nucleons:
\beqn\label{mav}
c_\pi=1+\frac{2m_\pi^2\alpha}{3F^2}\nonumber\\
c_N=1+\frac{9m_\pi^2g_A^2\alpha}{8F^2}
\eeqn

\section{Reaction Thresholds}

Knowing the LIV for nucleons, pions, photons and electrons, we proceed to study the reactions involved
in the GZK cutoff.
We follow the discussion in \cite{6,9}.

\subsection{Photo-Pion Production $\gamma + p \rightarrow p + \pi$}

Let us begin with the photo-pion production $\gamma + p
\rightarrow p + \pi$. Considering the corrections provided in the
dispersion relation (\ref{mav}) for pions and
nucleons, we note that, for the photo-pion production to proceed,
the following condition must be satisfied
\begin{eqnarray} \label{eq: pion}
2 \, \delta c \, E_{\pi}^{2} + 4E_{\pi} \omega \geq
\frac{m_{\pi}^{2}(2 m_{p}+m_{\pi})}{m_{p}+m_{\pi}}.
\end{eqnarray}
where $E_{\pi}$ is the produced pion energy and $\delta c =
c_{p} - c_{\pi}$.

\subsection{Pair Creation $\gamma + p \rightarrow p + e^{+} + e^{-}$}

Pair creation, $\gamma + p \rightarrow p + e^{+} + e^{-}$, is
greatly abundant in the sector previous to the GZK limit. When the
dispersion relations for fermions are considered for both protons
and electrons, it is possible to find
\begin{eqnarray} \label{pair creation}
\delta c \frac{m_{e}}{m_{p}} E^{2} + E \omega \geq m_{e}
(m_{p} + m_{e}), \label{eq: pair}
\end{eqnarray}
where $E$ is the incident proton energy and $\delta c =
c_{p} - c_{e}$.

\section{Bounds}

In order to study the threshold conditions (\ref{eq: pion})
and (\ref{eq: pair}), in the context of the GZK
anomaly, we must establish some criteria. Firstly, as it is seen
in \cite{Berezinsky,9}, the conventionally obtained theoretical
spectrum provides a very good description of the phenomena up to
an energy $\sim 4 \times 10^{19}$ eV. The main reaction taking
place in this well described region is the pair creation $\gamma +
p \rightarrow p + e^{+} + e^{-}$ and, therefore, no modifications
are present for this reaction up to $\sim 4 \times 10^{19}$ eV. As
a consequence, and since threshold conditions offer a measure of
how modified kinematics is, we will require that the threshold
condition (\ref{eq: pair}) for pair creation not be substantially
altered by the new corrective terms.

Secondly, we have the GZK anomaly itself, which we want
to explain. Since for energies greater than $\sim 8 \times
10^{19}$ eV the conventional theoretical spectrum does not fit the
experimental data well, we shall require that QG corrections be
able to offer a violation of the GZK-cutoff. The dominant reaction
in the violated $E > 8 \times 10^{19}$ region is the photo-pion
production and, therefore, we further require that the new
corrective terms present in the kinematical calculations be able
to shift the threshold significantly to preclude the reaction.

We begin our analysis with the threshold condition for pair
production. In this case we have:
\begin{eqnarray}
\delta c \, \frac{m_{e}}{m_{p}} E^{2} + E \omega \geq m_{e}
(m_{p} + m_{e}),
\end{eqnarray}
with $\delta c = c_{p} - c_{e}$. As is clear from
the above condition, the minimum soft-photon energy
$\omega_{\mathrm{min}}$ for the pair production to occur, is
\begin{eqnarray}
\omega_{\mathrm{min}} = \frac{m_{e}}{E} (m_{p} + m_{e}) - \delta
c \, \frac{m_{e}}{m_{p}} E.
\end{eqnarray}
It follows therefore that the condition for a significant increase
or decrease in the threshold energy for pair production becomes
$|\delta c| \geq m_{p} (m_{p} + m_{e}) / E^{2}$. In this way,
if we do not want kinematics to be modified up to a reference
energy $E_{\mathrm{ref}} = 3 \times 10^{19}$, we must impose the
following constraint
\begin{eqnarray} \label{eq: bound alpha}
|c_{p} - c_{e}| < \frac{(m_{p} + m_{e} ) m_{p} }{
E_{\mathrm{ref}}^{2}} = 9.8 \times 10^{-22}.
\end{eqnarray}
Similar treatments can be found for the analysis of other
astrophysical signals like the Mkn 501 $\gamma$-rays \cite{Stecker
& Glashow}, when the absence of anomalies is considered.

Let us now consider the threshold condition for the photo-pion
production. We have
\begin{eqnarray} \label{eq: pion2}
2 \, \delta c \, E_{\pi}^{2} + 4E_{\pi} \omega \geq
\frac{m_{\pi}^{2}(2 m_{p}+m_{\pi})}{m_{p}+m_{\pi}}.
\end{eqnarray}
It is possible to find that for the above condition to be violated
for all energies $E_{\pi}$ of the emerging pion, and therefore no
reaction to take place, the following inequality must hold
\begin{eqnarray} \label{eq: pi-p}
c_{\pi} - c_{p} > \frac{2 \omega^{2} (m_{p} +
m_{\pi})}{m_{\pi}^{2} (2 m_{p} + m_{\pi})} = 3.3 \times 10^{-24}
\left[ \omega / \omega_{0} \right] ^{2}.
\end{eqnarray}
where $\omega_{0} = K T = 2.35 \times 10^{-4}$ eV is the thermal
CMBR energy. 

Combining the two reactions and the standard values, $m_\pi=139 Mev,g_A=1.26, F=92.4 Mev$,
we get an upper and lower  bound on $\alpha$

\be
2.2\times 10^{-21} > -\alpha > 1.3 \times 10^{-24}\label{bound}
\ee

First of all, we notice that $\alpha < 0$, in order to suppress the photopion production, thus
removing the GZK cutoff. This implies that photons are the fastest particles and they arrive before
neutrinos coming from the same source of GRB. Moreover, photons become unstable. They decay in a electron 
positron pair above an energy $E_0$\cite{6}. See below.

Since $c_{photon} > c_{proton}$, the strong bound of \cite{cg2} is avoided: Proton is stable under
Cerenkov radiation in vacuum.

If no GZK anomaly is confirmed in future experimental
observations, then we should state a stronger bound for the
difference $c_{\pi} - c_{p}$. Using the same assumptions
to set the restriction (\ref{eq: bound alpha}) when the primordial
proton reference energy is  $E_{\mathrm{ref}} = 2 \times 10^{20}$
eV, it is possible to find
\begin{eqnarray}
|c_{\pi} - c_{p}| < 2.3 \times 10^{-23}.
\end{eqnarray}
In terms of $\alpha$, this last bound may
be read as
\begin{eqnarray}
|\alpha|< 9.1\times 10^{-24},
\end{eqnarray}
which is a stronger bound over $\alpha$ than (\ref{eq: bound
alpha}), offered by pair creation.

{\bf Photon unstability}

It has been pointed out in \cite{cg2,6} that if $c_{photon}>c_{electron}$ then the process
$\gamma  \rightarrow  e^{+} + e^{-}$ is allowed above an energy $E_0$:
\be
E_0=m_e\sqrt{\frac{2}{\delta c}}
\ee
where $\delta c=c_\gamma-c_e$.

In our case, we have:
\beqn
\delta c_L=-\alpha(\frac{23}{6}e^2-(\frac{g^2}{cos^2\theta_w}(sin^2\theta_w-1/2)^2 +e^2+g^2/2)/2)\\
\delta c_R=-\alpha(\frac{23}{6}e^2-(e^2+\frac{g^2sin^4\theta_w}{cos^2\theta_w})/2)
\eeqn
Therefore, with 
\beqn\label{eepair}
EL_0=2.3\times 10^8 Gev\nonumber\\
ER_0=1.9\times 10^8 Gev
\eeqn

So, we should not detect photons with energies above $2.3\times 10^8 Gev$ 

{\bf Neutral pion Stability}

Following \cite{6} we study the main decay process of neutral pion 
$\pi_0 \rightarrow \gamma+\gamma$ .This becomes possible if $c_\gamma>c_\pi$
and above an energy
\be
E_{\pi}=\frac{m_{\pi}}{\sqrt{2(c_\gamma-c_\pi)}}
\ee
Using the bound  $c_\gamma-c_\pi<10^{-22}$ obtained in \cite{antonov}, we get
\be
|\alpha|<5.4\times 10^{-23}
\ee

In our numerical estimates we have chosen $\alpha=-5\times 10^{-23}$.

We get $E_{\pi}=10^{19} eV$. Therefore we expect that neutral pions above this energy are stable, so
they could be a primary component of UHECR. Photons will be unstable above this energy by the 
same mechanism. Notice however that photons are unstable at a lower energy due to electron-positron 
pair creation (\ref{eepair}).

\section{Conclusions}

In this paper we have computed the LIV induced by Quantum Gravity on Baryons and Mesons, using the Chiral Lagrangian
approach. This permitted to fix that $\alpha<0$, in order to explain the GZK anomaly. Studying several available processes,
we found bounds on $\alpha$:

From pair creation and absence of photopion creation:
$2.2\times 10^{-21} > -\alpha > 1.3 \times 10^{-24}$.

From pion stability and the most stringent experimental bound found in \cite{antonov}: $|\alpha|<5.4\times 10^{-23}$.

Then, several predictions are obtained:Photons are unstable above an energy $2.3\times 10^8 Gev$.

Neutral pions are stable
above an energy $E_{\pi}=10^{19} eV$; so they could be a primary component of UHECR, thus evading the GZK cutoff.

Moreover, in time of flight experiments, photons will arrive before neutrinos, assuming that they were emitted simultaneously
at the source. No energy dependence of the time delay should be observed.
The time delay during a flight
 from a source situated at a distance
 $D$ will be of the order of $(5\times 10^{-23}) D/c\sim 5\times 10^{-6}$ s, assuming $D=10^{10}$ light-years.


\section*{Acknowledgements}
 The work of JA is partially supported by Secretaria de Estado de Universidades e 
Investigaci\'on SAB2003-0238(Spain). He wants to thank A.A. Andrianov for several useful remarks; and interesting 
conversations with H.A. Morales-T\'ecotl, L.F.Urrutia, D. Sudarsky, C. Kounnas, C. Bachas, V. Kazakov, A. Bilal, 
M.B. Gavela, E. Alvarez, A. Gonz\'alez-Arroyo and G. Palma. He acknowledges the hospitality of the Perimeter Institute and 
Ecole Normale Superieure,Paris.


\end{document}